\documentclass[preprint2]{aastex}
\usepackage{txfonts}
\usepackage{epsfig}
\usepackage{graphicx}
\usepackage{natbib}

\shorttitle{Is the current lack of solar activity only skin deep?}
\shortauthors{Broomhall et al.}
\begin{document}

\title{Is the current lack of solar activity only skin deep?}

\author{A.-M. Broomhall$^1$, W.~J. Chaplin$^1$, Y. Elsworth$^1$, S.~T. Fletcher$^2$, and R. New$^2$}
\affil{$^1$School of Physics and Astronomy, University of
Birmingham, Edgbaston, Birmingham B15 2TT, UK}
\email{amb@bison.ph.bham.ac.uk, wjc@bison.ph.bham.ac.uk,
ype@bison.ph.bham.ac.uk}\affil{$^2$Faculty of Arts, Computing,
Engineering and Sciences, Sheffield Hallam University, Sheffield S1
1WB, UK} \email{S.Fletcher@shu.ac.uk, R.New@shu.ac.uk}

\begin{abstract} The Sun is a variable star whose magnetic activity
and total irradiance vary on a timescale of approximately 11 years.
The current activity minimum has attracted considerable interest
because of its unusual duration and depth. This raises the question:
what might be happening beneath the surface where the magnetic
activity ultimately originates? The surface activity can be linked
to the conditions in the solar interior by the observation and
analysis of the frequencies of the Sun's natural seismic modes of
oscillation - the $p$ modes. These seismic frequencies respond to
changes in activity and are probes of conditions within the Sun. The
Birmingham Solar-Oscillations Network (BiSON) has made measurements
of $p$-mode frequencies over the last three solar activity cycles,
and so is in a unique position to explore the current unusual and
extended solar minimum. We show that the BiSON data reveal
significant variations of the $p$-mode frequencies during the
current minimum. This is in marked contrast to the surface activity
observations, which show little variation over the same period. The
level of the minimum is significantly deeper in the $p$-mode
frequencies than in the surface observations. We observe a
quasi-biennial signal in the $p$-mode frequencies, which has not
previously been observed at mid- and low-activity levels. The stark
differences in the behavior of the frequencies and the surface
activity measures point to activity-related processes occurring in
the solar interior, which are yet to reach the surface, where they
may be attenuated.
\end{abstract}

\keywords{methods: data analysis - Sun: activity - Sun:
helioseismology - Sun: oscillations}

\section{Introduction}
The level of the Sun's magnetic activity is observed to vary on an
11-year time scale and we are currently in the minimum between
cycles 23 and 24. The current solar minimum is attracting a great
deal of attention as it is proving to be quite unusual. Observations
of surface and atmospheric effects, such as the number of visible
sunspots, the rate of occurrence of solar flares and the strength of
the solar wind, highlight just how quiet the Sun
is\footnote[3]{http://science.nasa.gov/headlines/y2009/01apr\_deepsolarminimum.htm}.

The Sun's activity cycle influences everyday life on the Earth. The
rate of occurrence of solar flares is dependent on the number of
spots on the Sun's surface, and large solar flares can disrupt
satellite communications and cause power outages. Coronal mass
ejections (CMEs), which are another source of radiation that can
disrupt life on the Earth, are still being observed regularly on the
Sun, even in this unusual solar minimum (see the STEREO COR1 CME
catalog\footnote[4]{
http://cor1.gsfc.nasa.gov/docs/prelim\_events/COR1prelimCMErate\_Feb2009.pdf}).
Cosmic rays, which are a significant space radiation hazard, are
anticorrelated with the solar cycle. The level of solar activity is
possibly correlated to the Earth's climate \citep[see, for
example,][and references therein]{Lockwood2007}. Space-weather
groups use predictions of solar cycles to anticipate the amount of
orbital drag experienced by satellites.

The next solar cycle has already proven difficult to predict as the
current solar minimum is lasting significantly longer than expected.
\citet{Pesnell2008} reviews 50 predictions for the size and timing
of cycle 24 and finds a wide range of results, especially in
comparison to predictions made before the previous solar cycle
\citep{Joselyn1997}. For example, predictions of when cycle 24 will
reach its maximum range from 2009 December \citep{Maris2006} to 2014
December \citep{Tsirulnik1997}. In fact, the official NOAA, NASA and
ISES Solar Cycle 24 Prediction Panel, which studied the predictions
collated by \citet{Pesnell2008}, failed to reach a consensus on when
the peak of cycle 24 will occur and how active the upcoming cycle
will be. Meanwhile, the number of predictions for cycle 24 is ever
increasing.

Surface measures of the Sun's activity, such as the number of
sunspots, which are used to aid cycle predictions, indicate that we
are still in an extended solar-cycle minimum. We ask the question:
can we learn anything about this unusual solar minimum from the
Sun's interior?

To answer this question we investigate the variation with the solar
cycle of the frequencies of the Sun's natural resonant oscillations,
which are known as $p$ modes. Solar $p$ modes are trapped in
cavities below the surface of the Sun and their frequencies are
sensitive to properties, such as temperature and mean molecular
weight, of the solar material. It has been known since the mid 1980s
\citep{Woodard1985} that $p$-mode frequencies vary throughout the
solar cycle with the frequencies being at their largest when the
solar activity is at its maximum. By examining the changes in the
observed $p$-mode frequencies throughout the solar cycle, we can
learn about solar-cycle-related processes that occur beneath the
Sun's surface.

The Birmingham Solar-Oscillations Network
\citep[BiSON;][]{Chaplin1996} makes Sun-as-a-star (unresolved)
Doppler velocity observations, which are sensitive to the $p$ modes
with the largest horizontal scales (or the lowest angular degrees,
$\ell$). Consequently, the frequencies measured by BiSON are of the
truly global modes of the Sun. These modes travel to the Sun's core
but their dwell time at the surface is longer than at the solar core
because the sound speed inside the Sun increases with depth.
Consequently, the low-$\ell$ modes are most sensitive to variations
in regions of the interior that are close to the surface and so are
able to give a global picture of the influence of near-surface
activity. BiSON is a network of autonomous ground-based
observatories that are strategically positioned at various
longitudes in order to provide as continuous coverage as possible of
the Sun. BiSON is in a unique position to study the changes in
oscillation frequencies that accompany the solar cycle as it has now
been collecting data for over 30 years. However, when the network
was first established the quality of the data was relatively poor,
in comparison to recent years, because of the sporadic coverage of
the observations. Here, we have been able to analyze the mode
frequencies observed during the last two solar cycles in their
entirety.

\section{Analysis and results}

The precision with which $p$-mode frequencies can be determined is
directly related to the length of time series under consideration.
Consequently, $p$-mode frequencies are often determined from time
series whose lengths are of the order of years. However, a
compromise must be made here regarding the appropriate length of
time series for study so that changes in the solar cycle can be
resolved. Before 1985 April the observed data are sparse because the
early BiSON data were collected from just one or two sites and,
initially, just in summer months in the northern hemisphere.
Therefore, when considering the data collected before 1985 April the
$p$-mode frequencies were obtained from seven time series of
different lengths, which reflected the availability of data. These
data are included here for completeness but are not used in the
later analysis as the data are too sparse to provide reliable
results. After the third BiSON site, at Carnarvon, Western
Australia, was established in 1985 the duty cycle of the BiSON data
increased significantly. Therefore, after 1985 April 22 the time
series were truncated to 54 days in length, which corresponds to
approximately twice the rotation period of the solar surface. The
analysis performed here concentrates on the data obtained after this
date.

A standard likelihood maximization method was used to fit the power
spectra of these time series \citep[e.g.,][]{Chaplin1999}, enabling
the observed mode frequencies to be determined. We have concentrated
on the strongest modes of oscillation, which are in the frequency
band from 2100 to 3500$\,\rm\mu Hz$.

\begin{figure*}
  \centering
  \includegraphics[clip, width=0.68\textwidth]{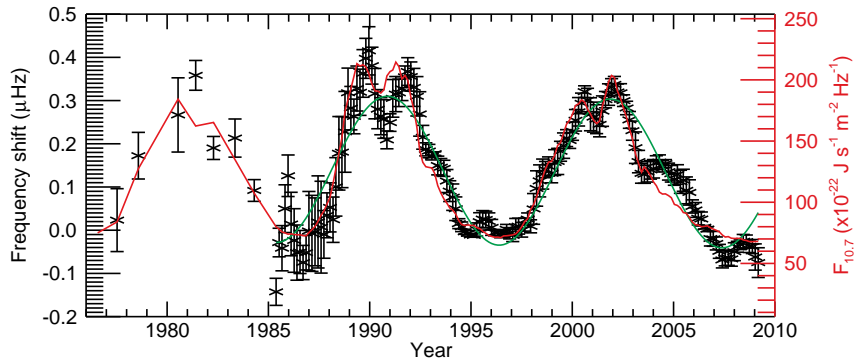}
  \caption{Smoothed frequency shifts observed in BiSON data during cycles 21, 22, and 23.
  The error bars associated with the frequency-shift data are due
  to uncertainties in the fitted oscillation frequencies. The green
  line is a sinusoid which has been fitted to the seismic data. Overplotted
  in red, and using the right-hand axis, is the $\rm F_{10.7}$,
  which has been scaled using a linear fit to the observed frequency shifts.}\label{figure[flux shifts]}
\end{figure*}

To maintain consistency with previous work
\citep[e.g.,][]{Chaplin2007} a minimum activity reference set was
determined by averaging the frequencies from time series observed
during the minimum activity epoch at the boundary between cycle 22
and cycle 23. We then defined the solar-cycle frequency shifts,
$\delta\nu_{n,l}(t)$, as the differences between frequencies given
in the minimum activity reference set and the frequencies of the
corresponding modes observed at different epochs and, consequently,
different levels of activity. The size of a frequency shift has a
well-known dependency on frequency and mode inertia and these
dependencies were removed in the manner described in
\citet{Chaplin2007}. This allowed a weighted average of the
frequency shifts observed for each time series to be determined,
which provided a mean frequency shift for each epoch,
$\delta\nu(t)$. Figure \ref{figure[flux shifts]} shows the
frequency-shift data that were observed during cycles 21, 22, and
23. The data are shown up to 2009 April. After 1985 April the data
plotted in Figure \ref{figure[flux shifts]} have been smoothed over
five points, as the mean frequency shifts were obtained from time
series of the same length.

One of the most obvious manifestations of the Sun's solar cycle is
the variation in the number of sunspots on the solar surface,
formally recorded as the International Sunspot Number (ISN).
However, the sunspot number is not the only variable that can be
used as a proxy for the solar activity. For example, the radio flux
emitted from the Sun at a wavelength of 10.7cm (hereafter $\rm
F_{10.7}$) also responds to changes in the solar cycle. Many authors
\citep[e.g.,][and references therein]{Elsworth1990, Howe2008} have
determined and commented upon the good correlations between shifts
in the $p$-mode frequencies and activity proxies such as the $\rm
F_{10.7}$ and the ISN. Therefore, for comparison purposes, the
average 10.7 cm flux has been plotted on top of the observed
frequency shifts in Figure \ref{figure[flux shifts]}. After the
first seven points plotted in Figure \ref{figure[flux shifts]} the
$\rm F_{10.7}$ has been smoothed as per the frequency shifts. To
determine the scale on which the proxy axis was plotted, a linear
least-squares fit between the observed frequency shifts and the
activity proxy was performed.

The rise and decline of cycle 23 appears to be slow in comparison to
cycle 22. The oscillation frequencies observed in the current
unusual solar minimum are significantly lower than the previous
minimum in 1996 and the most recent data indicate that we are still
on a downward trend. Notice that the shapes of the previous and
current solar minima are very similar, with both appearing to show a
double minimum. This implies that we may still have to wait some
time before the rising phase of cycle 24 begins. It is also
interesting to note that, if account is taken of the lower quality
of the data observed before 1990, the minimum between cycles 21 and
22 is also deeper than the last solar minimum. This could be because
both the minimum between cycles 21 and 22 and the current solar
minimum (between cycles 23 and 24) correspond to the same phase of
22-year magnetic Hale cycles.

It is clearly apparent from Figure \ref{figure[flux shifts]} that
there is an unusually large difference between the activity proxy
and the frequency shifts on the declining phase of cycle 23.
Notably, the frequency shifts are still observed to be changing
throughout the solar minimum, which is in contrast to the activity
proxy data that show little structure in its variation with time
since 2006.

\begin{figure*}
\centering
  \includegraphics[clip,
  width=0.68\textwidth]{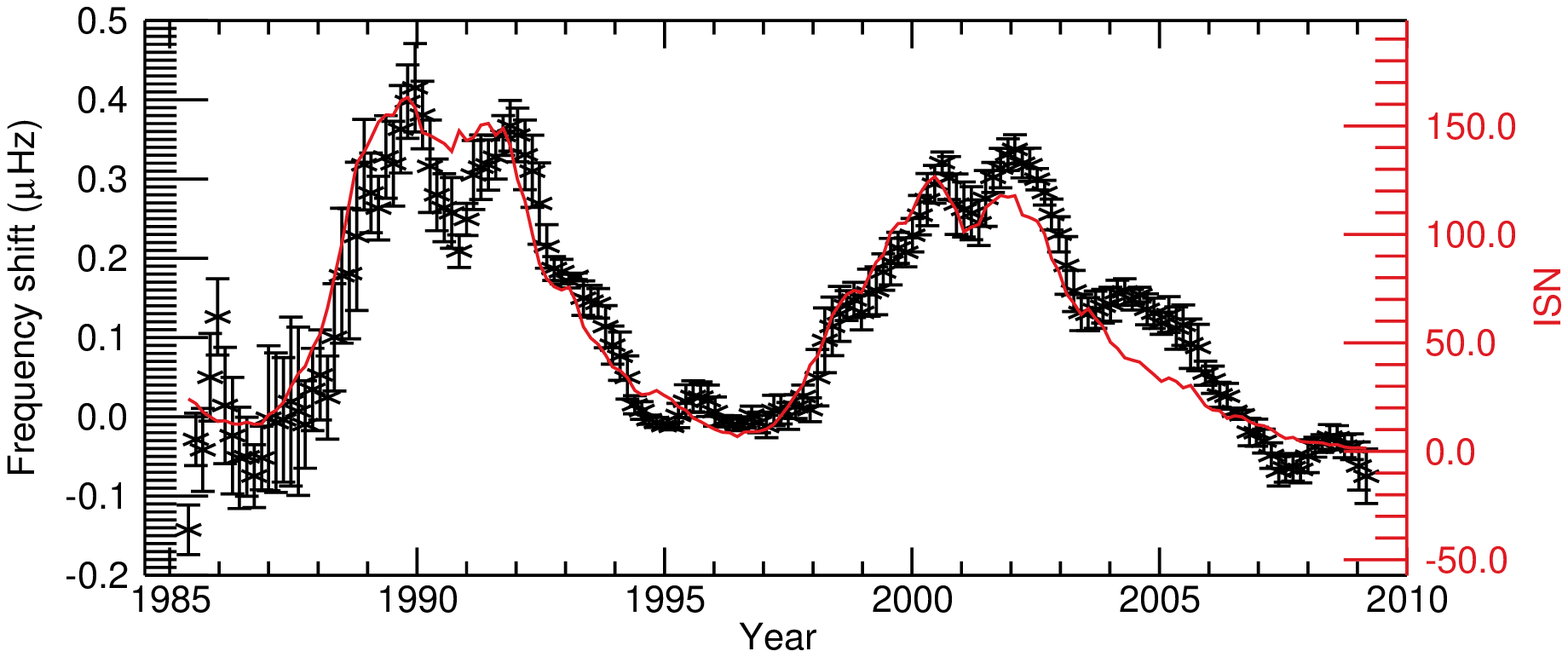}\\
  \includegraphics[clip,
  width=0.68\textwidth]{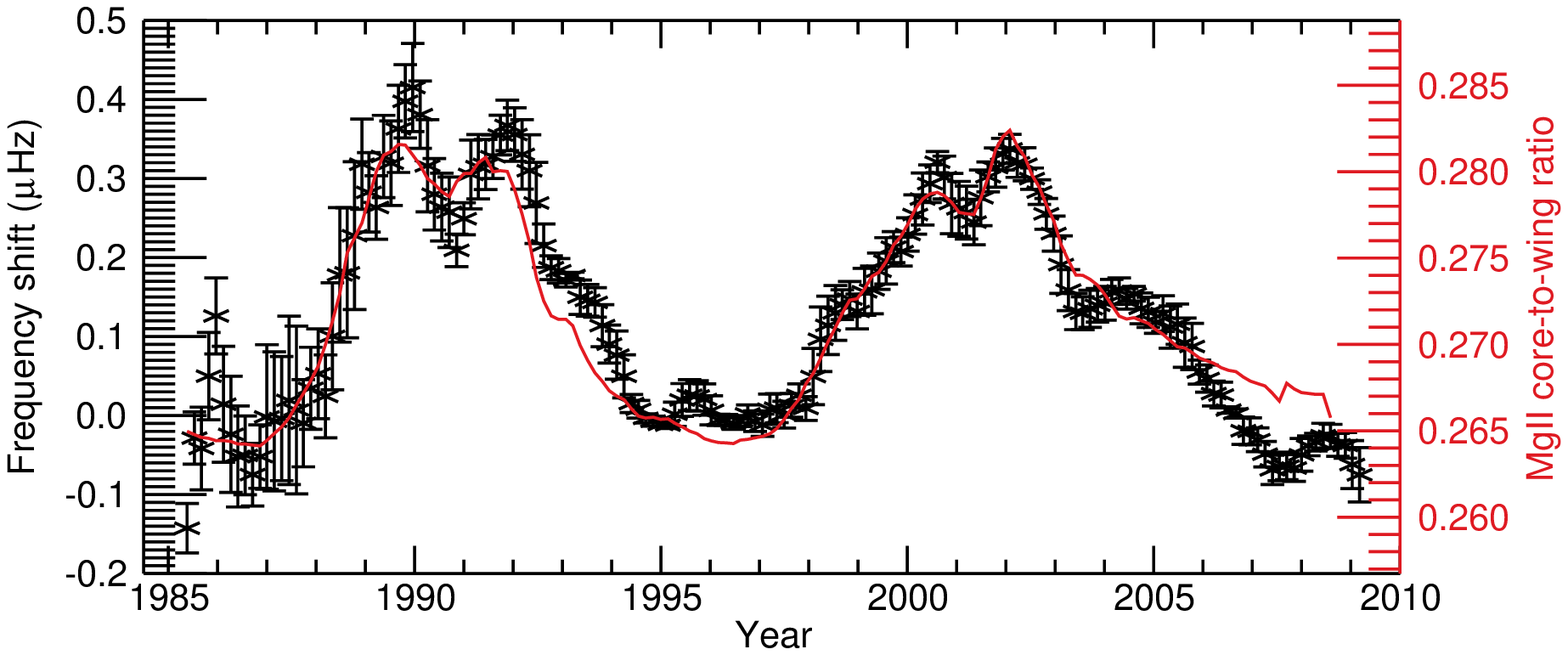}\\
  \caption{Comparison between the frequency shifts
  observed in the BiSON data and two different activity proxies.
  Top:
  overplotted in red is the ISN. Bottom: overplotted in red is
  the MgII core-to-wing ratio.}
\end{figure*}\label{figure[flux proxies]}

Figure \ref{figure[flux proxies]} allows a comparison between the
BiSON frequency shifts and two proxies of the Sun's activity, other
than the $\rm F_{10.7}$. As noted above the ISN is a measure of the
number of active sunspot groups and individual spots on the visible
disk. The MgII core-to-wing ratio \citep{Viereck2001} is determined
from space-based UV spectral irradiance measurements. The MgII H and
K line cores originate in the chromosphere and the ratio of the core
size to that of the more stable background gives a robust indication
of the chromospheric activity.

The $\rm F_{10.7}$ and the MgII core-to-wing ratio appear to show
better agreement with the frequency shifts than the ISN. This
behavior has been observed before \citep{Chaplin2007} and occurs
because the frequency shifts are sensitive to both the strong
\emph{and} weak components of the Sun's magnetic flux. The $\rm
F_{10.7}$ and the MgII core-to-wing ratio show similar relative
sensitivity to the different components of the Sun's magnetic flux,
whereas, the ISN is predominantly sensitive to the strong component
of the Sun's magnetic field.

For the purposes of this Letter it is most important to notice that
there are large discrepancies between, respectively, all three of
the proxies plotted in Figures \ref{figure[flux shifts]} and
\ref{figure[flux proxies]} and the frequency shifts in the declining
phase of cycle 23. Furthermore, the structure that is observed in
the frequency shifts in the current solar minimum is not replicated
by any of the proxies. It is also important to note that the seismic
minimum is decidedly lower than the minimum predicted by the three
proxies. The oscillations are sensitive to the conditions beneath
the solar surface, whereas the activity proxy is a measure of the
magnetic field that is present on the surface. We therefore assume
that there are processes happening in the solar interior, which have
yet to manifest themselves at the surface.

\subsection{Short-Term Variations in the Frequency Shifts}
We now turn our attention to the shorter-term variable structure in
the frequency shifts visible on top of the general 11-year trend,
which has a period of around two years. Similar ``quasi-biennial''
variation of activity proxies has been noted before in data from the
green coronal emission line at 530.3\,nm at high solar activity
\citep{Vecchio2008}. These authors used variations near the solar
equator and poles and evidence was found of quasi-biennial
variability at both low and high latitudes. An explanation of such
quasi-biennial behavior has been put forward in terms of two
different types of dynamo operating at different depths
\citep{Benevolonskaya1998a, Benevolonskaya1998b}. Elsewhere
\citep{Saar2002}, it has been possible to identify shorter secondary
periods in the activity cycles of some stars and it is conceivable
that asteroseismology may be able to detect such effects in data
sets which will become available, for example, from the recently
launched $Kepler$ satellite \citep{Christensen2008}.

In spite of the known behavior of the proxies, this is the first
time that the quasi-biennial variability has been noted in seismic
data. Furthermore, the signature of the two-year signal in the
activity index was previously restricted to times of moderate to
high solar activity. For the first time, we are seeing this
short-term variability at low solar activity. To investigate this
further, a sine wave was fitted to the frequency shifts observed
after 1985 April (see Figure \ref{figure[flux shifts]}). The sine
wave took the form
\begin{equation}\label{equation[sine wave]}
    S(\delta t)=A_0\sin\left(2\pi
    A_1\delta t+A_2\right)+A_3\delta t+A_4,
\end{equation}
where $\delta t$ is the time in years after 1985 April 22, which is
the start date of the first 54\,d time series. The $A_i$
coefficients are given in Table \ref{table[A coefficients]}. The
sine wave was calculated using a weighted least-squares fit to the
unsmoothed frequency shifts. In equation \ref{equation[sine wave]},
the first collection of terms accounts for the sine wave structure
of the frequency shifts. The second and third terms account for an
offset which linearly decreases with time.

\begin{table}
  \centering
  \caption{The $A_i$ coefficients used in calculating equation \ref{equation[sine wave]}.}
  \begin{tabular}{cc}
    \hline
    \hline
    $\qquad i\qquad$ & $A_i$ \\
    \hline
    0 & $0.17\,\rm \mu Hz$ \\
    1 & $0.091\,\rm yr^{-1}$ \\
    2 & $4.7 \,\rm rad$ \\
    3 & $-0.00057\,\rm \mu Hz\,yr^{-1}$ \\
    4 & $0.14\,\rm \mu Hz$ \\
    \hline
  \end{tabular}
  \label{table[A coefficients]}
\end{table}

\begin{figure*}
  \centering
  \includegraphics[clip, width=0.68\textwidth]{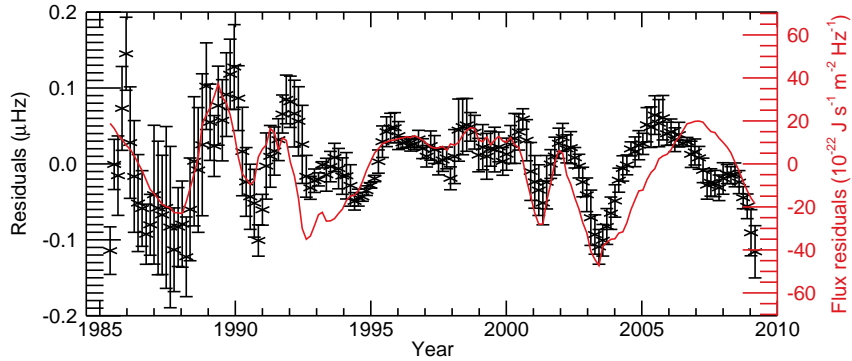}
  \caption{Residuals observed between the best-fitting sine wave described by
  equation \ref{equation[sine wave]} and the observed frequency
  shifts. Overplotted in red, and using the right-hand axis, are the
  residuals between the sine wave, whose amplitude was scaled so as
  to be appropriate to the flux measurements, and the $\rm
  F_{10.7}$. The $\rm
  F_{10.7}$ residuals have been scaled using a linear fit to the observed frequency shifts.}\label{figure[residuals]}
\end{figure*}

We have determined the residuals between the best-fitting sine wave
and the observed frequency shifts and the results are plotted in
Figure \ref{figure[residuals]}. Flux residuals were also determined
between a sine wave, which was scaled in amplitude using the linear
fit between the flux and the frequency shifts, and the $\rm
F_{10.7}$. Clearly, there is a lot of structure in both sets of
observed residuals and since $\sim2002$ the frequency-shift
residuals are quite substantial in size. Large discrepancies between
the two sets of residuals are clearly evident. A periodogram of the
data from the last two cycles shows significant periods of about two
and three years with greater than 99\% confidence, but the phase of
the signal is not locked to that of the underlying 11-year cycle and
several incidences of phase jumps are evident, meaning that periods
are not easy to determine. The three-year periodicity is not visible
in a periodogram of the activity proxy residuals.

\section{Summary}

The solar-cycle shifts that are observed in $p$-mode frequencies are
usually well correlated with proxies of the Sun's activity, such as
the 10.7 cm radio flux. However, in the declining phase of cycle 23
and the current solar minimum we find unusually large differences
between the frequencies observed in BiSON data and the activity
levels. The current cycle minimum indicated by the helioseismic data
is significantly deeper than the minima observed by the activity
proxies and the structure that is clearly evident in the frequency
shifts is not replicated in the proxy data. We also observe a
quasi-biennial signal in the $p$-mode frequencies at \emph{all}
activity levels. Interestingly, this signal is \emph{only} visible
at \emph{high}-activity levels in the proxy data. As the frequency
shifts respond to conditions beneath the surface of the Sun whereas
the proxies respond to changes at or above the surface, we suggest
that these differences may be caused by changes in the magnetic flux
that have yet to manifest at the surface. It is also possible that
the magnetic flux responsible for the discrepancies between the
frequency shifts and the activity proxies will never reach the Sun's
surface. The analysis presented here was based on averages made over
groups of modes ($0\le\ell\le3$, $14\le n\le 23$). Further work on
individual modes may allow one to isolate the location of the
variability because each mode shows a different sensitivity to the
latitudinal distribution of the surface activity. Such an analysis
is currently in progress.

In the seismic data, the previous solar minimum exhibited a double
minimum and it appears that the current solar minimum is showing a
similar structure. Furthermore, the most recent $p$-mode frequencies
indicate that the current minimum is still declining. Therefore, it
may still be some time before we observe the rising phase of cycle
24. There have been suggestions that this recent strange behavior of
the Sun is indicative that the current grand maximum is about to end
\citep{Abreu2008}. That would indeed be an occurrence of great
significance. Although our results cannot predict whether this is
true they do indicate that the next solar cycle should be observed
with great interest.

\acknowledgements This Letter utilizes data collected by the
Birmingham Solar-Oscillations Network (BiSON). We thank the members
of the BiSON team, both past and present, for their technical and
analytical support. We also thank P. Whitelock and P. Fourie at the
South African Astronomical Observatory (SAAO), the Carnegie
Institution of Washington, the Australia Telescope National Facility
(Australian Commonwealth Scientific and Research Organization,
CSIRO), E.J. Rhodes (Mt. Wilson, Californa) and members (past and
present) of the Instituto de Astrofisica de Canarias (IAC),
Tenerife. BiSON is funded by the Science and Technology Facilities
Council (STFC). The authors also acknowledge the financial support
of STFC.

\end{document}